\documentclass[letter]{sig-alternate-05-2015}
\mathchardef\Gamma="0100 \mathchardef\Delta="0101
\mathchardef\Theta="0102 \mathchardef\Lambda="0103
\mathchardef\Xi="0104 \mathchardef\Pi="0105
\mathchardef\Sigma="0106 \mathchardef\Upsilon="0107
\mathchardef\Phi="0108 \mathchardef\Psi="0109
\mathchardef\Omega="010A

\newcommand{\outline}[1]{}

\usepackage{cite}
\usepackage{xspace}
\usepackage{url}
\usepackage{graphicx}
\usepackage{latexsym}
\usepackage{amssymb}
\usepackage{amsfonts}
\usepackage{psfrag}
\usepackage{wrapfig}
\usepackage{comment}
\usepackage{alltt}
\usepackage{color}

\newcommand{\ie}{\emph{i.e.}\xspace}
\newcommand{\eg}{\emph{e.g.}\xspace}

\newcommand{\Comment}[1]{}

\newcommand{\cut}[1]{}
\usepackage{subfigure,algorithm,algorithmic}
\usepackage{bbm}
\usepackage[table,xcdraw]{xcolor}
\usepackage{pifont}
\usepackage{longtable}
\usepackage{url}
\usepackage{balance}
\usepackage{epstopdf}
\usepackage{color, colortbl}
\definecolor{darkgreen}{RGB}{0,90,90}
\usepackage{graphicx}
\usepackage{multirow}
\usepackage{booktabs}
\usepackage{pdftexcmds}
\usepackage[export]{adjustbox}
\usepackage{longtable}
\newcommand{\presec}{\vspace{-0.0in}}
\newcommand{\postsec}{\vspace{-0.0in}}

\usepackage{listings}
\usepackage{color}
\definecolor{lightgray}{rgb}{.9,.9,.9}
\definecolor{darkgray}{rgb}{.4,.4,.4}
\definecolor{purple}{rgb}{0.65, 0.12, 0.82}

\lstdefinelanguage{JavaScript}{
  keywords={typeof, new, true, false, catch, function, return, null, catch, switch, var, if, in, while, do, else, case, break},
  keywordstyle=\color{blue}\bfseries,
  ndkeywords={class, export, boolean, throw, implements, import, this},
  ndkeywordstyle=\color{darkgray}\bfseries,
  identifierstyle=\color{black},
  sensitive=false,
  comment=[l]{//},
  morecomment=[s]{/*}{*/},
  commentstyle=\color{purple}\ttfamily,
  stringstyle=\color{red}\ttfamily,
  morestring=[b]',
  morestring=[b]"
}

\begin{document}

\title{A First Look at Ad-block Detection -- \\A New Arms Race on the Web}
\author{Muhammad Haris Mughees$^1$, Zhiyun Qian$^2$, Zubair Shafiq$^3$, Karishma Dash$^2$, Pan Hui$^1$\\
\affaddr{Hong Kong University of Science and Technology, University of California-Riverside, The University of Iowa}
}
\maketitle
\begin{abstract}
The rise of ad-blockers is viewed as an economic threat by online publishers,
especially those who primarily rely on advertising to support their services.
To address this threat, publishers have started retaliating by employing ad-block detectors,
which scout for ad-blocker users and react to them by
restricting their content access and pushing them to whitelist the website or disabling
ad-blockers altogether.
The clash between ad-blockers and ad-block detectors has resulted in a new arms race on the web.

In this paper, we present the first systematic measurement and analysis of ad-block detection on the web.
We have designed and implemented a machine learning based technique to automatically detect ad-block detection,
and use it to study the deployment of ad-block detectors on Alexa top-100K websites.
The approach is promising with precision of 94.8\% and recall of 93.1\%.
We characterize the spectrum of different strategies used by websites for ad-block detection.
We find that most of publishers use fairly simple passive approaches for ad-block detection.
However, we also note that a few websites use third-party services, \eg PageFair, for ad-block detection and response.
The third-party services use active deception and other sophisticated tactics to detect ad-blockers.
We also find that the third-party services can successfully circumvent ad-blockers and display ads on publisher websites.
\end{abstract}

\section{Introduction}
The online advertising industry has been largely fueling the World Wide Web for the past many years.
According to the Interactive Advertising Bureau (IAB), the annual online ad revenues for 2014 totaled \$49.5 billion in 2014, which is 15.6\% higher than in 2013 \cite{iabadreport}.
Online advertising plays a critical role in allowing web content to be offered free of charge to end-users, with the implicit assumption that end-users agree to watch ads to support these ``free'' services.
However, online advertising is not without its problems.
The economic magnetism of online advertising industry has made ads an attractive target for various types of abuses, which are driven by incentives for higher monetary benefits.
Since publishers are paid on a per-impression or per-click basis, many publishers choose to place ads such that they interfere with the organic content and cause \emph{annoyance} to end-users \cite{goldstein2013cost}.
They include anything from autoplay video ads, rollovers, pop-ups, and flash animation ads to the ever-popular homepage takeover with sidebars that follow user scrolling.
Another major issue with online advertising is the widespread tracking of users across websites raising \emph{privacy} and \emph{corporate surveillance} concerns.
Several recent studies have shown that ad exchanges aggressively profile users and invade user privacy \cite{guha2011privad}.
\emph{Malvertising} (using ads to spread malware) is also on the rise \cite{li2012knowing,zarras14malads}.

In addition to the above problems, many users simply desire an ad-free web experience which is much cleaner and smoother.
Therefore, ad-blockers have become popular in recent years and they can block ads seamlessly without requiring any user input.
A wide range of ad-blocking extensions are available for popular web browsers such as Chrome and Firefox \cite{bestadblockers}.
Adblock Plus is most prominent among all these extensions \cite{adp}.
According to a recent academic study, 22\% of the most active residential broadband users of a major European ISP use Adblock Plus \cite{pujol15adblockinwild}.
In addition, it is estimated in a recent report~\cite{pagefairreport} that \$22 billion will be lost due to ad-blocking in 2015, almost twice the amount estimated in 2014.
To the advertisement industry and content publishers, ad-blockers are becoming a growing threat to their business model.
To combat this, two strategies have emerged: (1) companies such as Google and Microsoft have begun to pay ad-blockers to have their ads whitelisted; and (2) websites have begun to detect the presence of ad-blockers and may refuse to serve any user with ad-blocker turned on, \eg Yahoo mail reportedly did so recently~\cite{yahoo_mail_incident}.

As not every website is willing or capable of paying ad-blockers, the 2nd strategy becomes a low-cost solution that can be easily deployed.
Even though anecdotes exist about websites starting to detect ad-blockers, the scale at which this occurs remains largely unknown.
To fill this gap, in this paper we perform the first systematic characterization of the ad-block detection phenomenon.
Specifically, we are interested in understanding:
(1) how many websites are performing ad-block detection;
(2) what type of technical approaches are used; and
(3) how can ad-blockers counter or circumvent such detection.

\noindent \textbf{Key Contributions.} The key contributions of the paper are the following:

\noindent\labelitemi\indent We conduct a measurement study of Alexa top-100K websites using a machine learning based approach to identify the websites that use ad-block detection.
The approach is promising with precision of 94.8\% and recall of 93.1\%.
The results show that around 300--1100 websites are currently performing ad-block detection (details in \S\ref{sec:measurement}).

\noindent\labelitemi\indent We cluster different ad-block detection approaches based on the JavaScripts that are inserted in the websites.
The results indicate that there is a spectrum of detection solutions ranging from fairly simple (passive detection) to complex (active deception).
We conduct several case studies to illustrate the strengths and limitations of different approaches.

\section{Background}  \label{sec:background}
In this section, we provide an overview of ad-blockers and ad-block detectors.

\vspace{0.05in} \noindent \textbf{The rise of ad-blockers.}
The issues with online ads has resulted in a proliferation of ad-blocking software.
Ad-blocking software (or ad-blocker) is an effective tool that blocks ads seamlessly, primarily
published as extensions in web browsers such as Chrome and Firefox \cite{bestadblockers}.
More recently, Apple has also allowed content blocking plugins for Safari on iOS devices \cite{appleadblock}.
Other popular relevant tools include Ghostery \cite{ghostery} and DisconnectMe \cite{diconnect};
however, they are primarily focused on protecting user privacy.
With respect to functionality, these ad-blockers
(1) block ads on websites and
(2) protect user privacy by filtering network requests that profile browsing behaviors.
Recent reports have shown that the number of users using ad-blocking software has rapidly increased worldwide.
According to PageFair, up to 198 million users around the world now use ad-blocking software \cite{pagefairreport}.
According a recent academic study, 22\% of the most active residential broadband users of a major European ISP use Adblock Plus \cite{pujol15adblockinwild}.
These ad-blocking users have been estimated to cost publishers more than \$22 billion in lost revenue in 2015 \cite{pagefairreport}.

\begin{figure}[t!]
\centering
\subfigure{
\includegraphics[width=1\columnwidth]{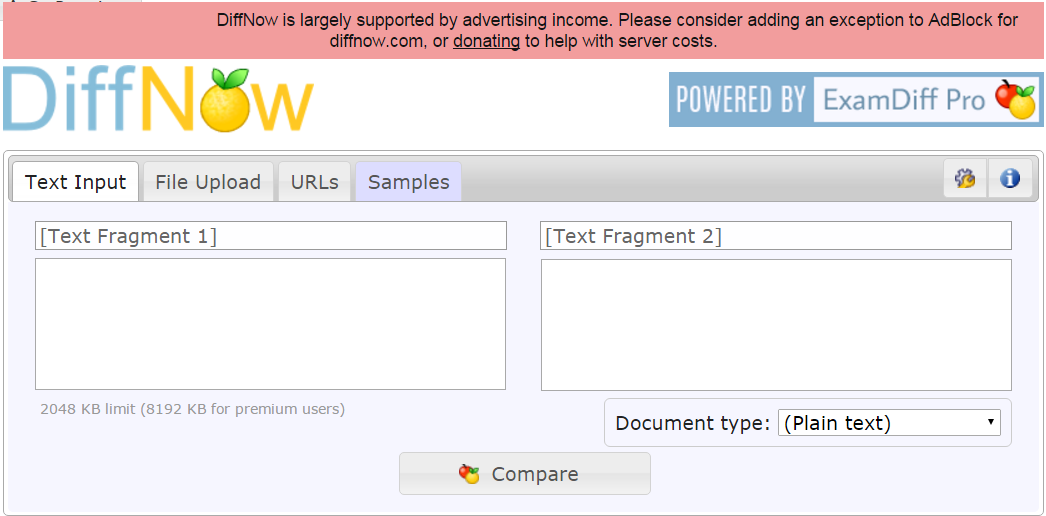}\label{fig:simple_adblock}
}
\subfigure{
\includegraphics[width=1\columnwidth]{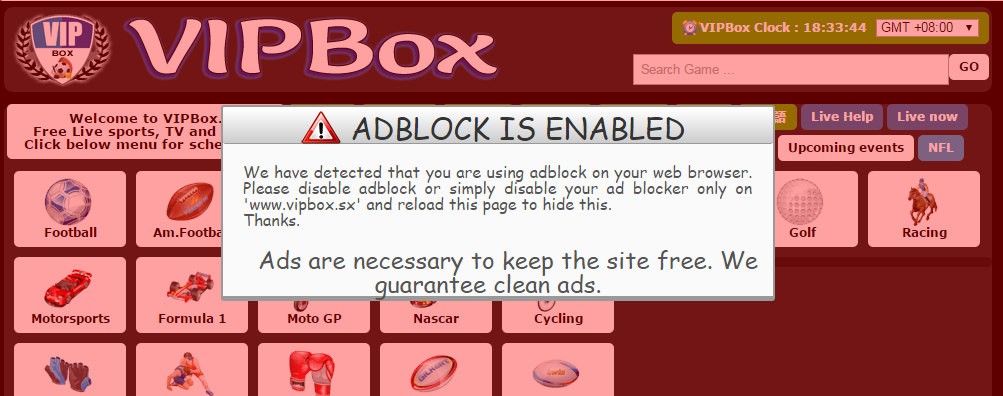}\label{fig:aggresive_adblock}
}
\vspace{-0.1in}
\caption{Typical ad-block detection responses}
\label{fig:sample_subfigures}
\vspace{-0.2in}
\end{figure}

\vspace{0.05in} \noindent \textbf{How do ad-blockers work?}
Ad-blockers eliminate ads by either \emph{page element removal} or \emph{web request blocking}.
For page element removal, ad-blockers use various CSS selectors to access the elements and remove them.
Similarly, for web requests, ad-blocker looks for particular URLs and remove the ones which belong to advertisers.
For both of these actions, ad-blockers are dependent on \emph{filter lists}
that contain the set of rules (as regular expressions) specifying the domains and element selectors to remove.
There are various kinds of filter lists available which can be included in ad-blockers.
Each of these lists serves a different purpose.
For example, Adblock Plus by default includes EasyList \cite{easylist}, which provides rules for removing ads from English websites.
Similarly, Fanboy \cite{fanboy} is another popular list that removes only annoying ads from websites.
Additionally, EasyPrivacy \cite{easyprivacy} helps ad-blockers to remove spy-wares.

\begin{figure*}[!t]
\centering
\subfigure{\includegraphics[width=0.3\textwidth]{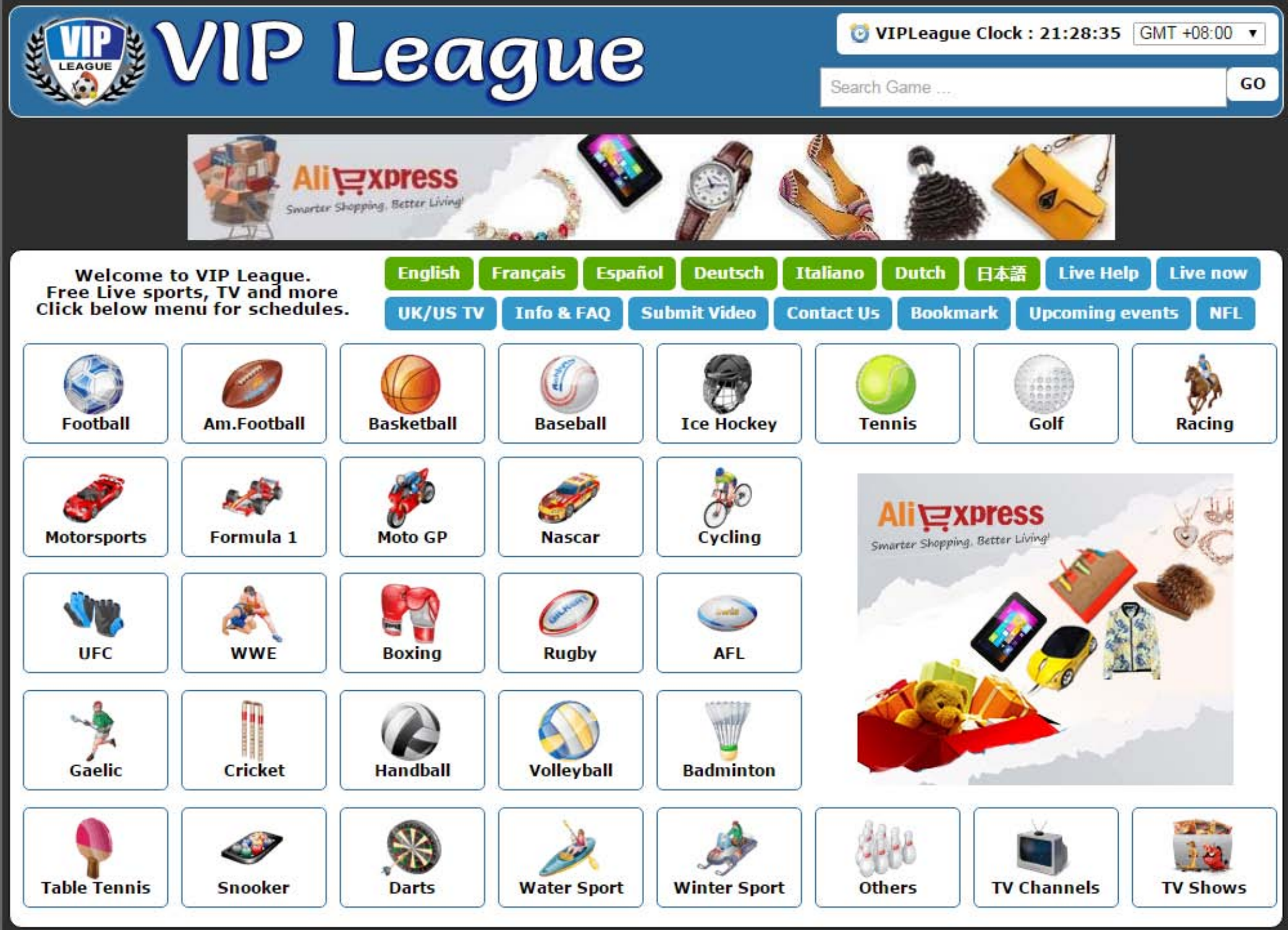}\label{fig:simple_adblock}}
\subfigure{\includegraphics[width=0.33\textwidth]{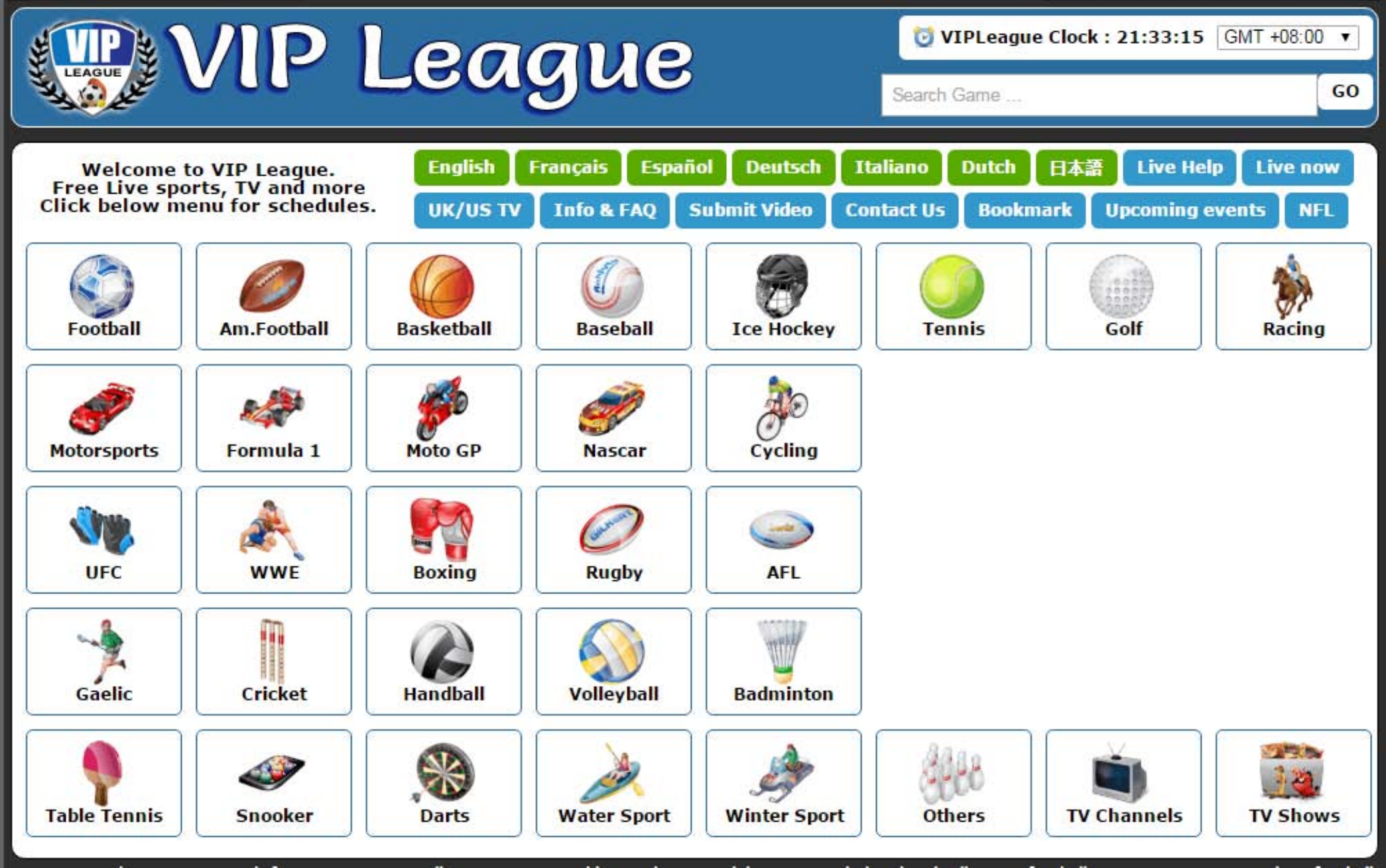}\label{fig:aggresive_adblock}}
\subfigure{\includegraphics[width=0.33\textwidth]{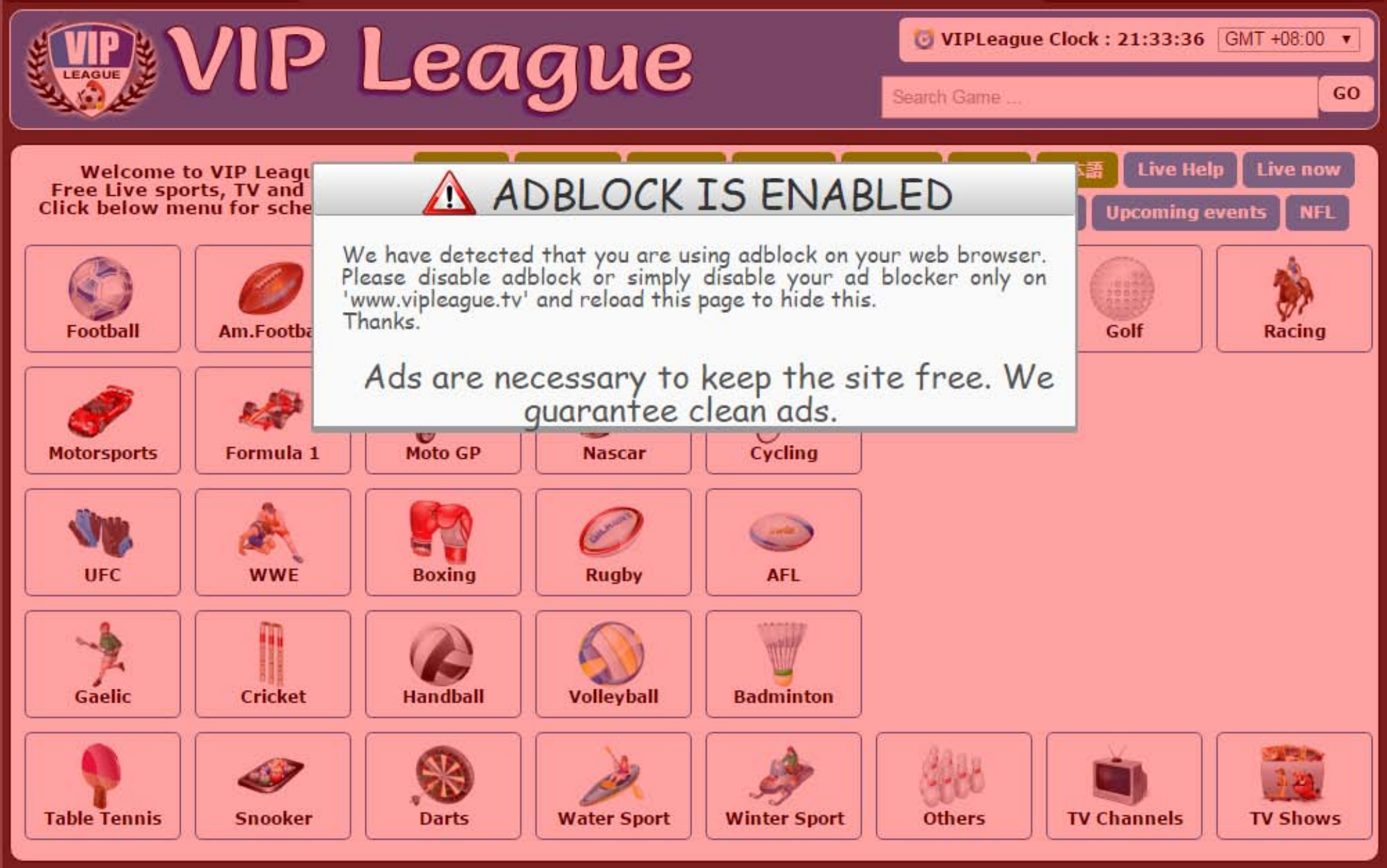}\label{fig:aggresive_adblock}}
\caption{Web page load evolution for \protect \url{http://www.vipleague.tv}. \textbf{Left:} The original website content is loaded. \textbf{Middle:} Ad-blocker removes ads from the page. \textbf{Right:} Ad-block detector blocks the content and shows a pop-up notification asking the user to disable ad-blocking software.}
\label{fig:adbtonoadb}
\end{figure*}

\vspace{0.05in} \noindent \textbf{The rise of ad-block detection.}
The widespread use of ad-blockers has prompted a cat-and-mouse game between publishers and ad-blocking software.
More specifically, publishers have started to detect whether users are visiting their websites while using ad-blocking software.
Once detected, publishers notify users to turn off their ad-blocking software.
These notifications can range from a mild non-intrusive message which is integrated inside website content to more aggressive blocking of website content and/or functionality.
Figure \ref{fig:sample_subfigures} shows examples of both cases.
We note that the aggressive approach refrains users from accessing any website content.
To detect the use of ad-blocking software, publishers include scripts in the code of their web pages.
When a user with the ad-blocking software opens such a website, these scripts typically monitor the visibility of ads on the page to identify the use of ad-blockers.
If ads are found hidden or removed by the scripts, publishers take countermeasures according to their policies.
It is noteworthy that the strategies used by publishers to detect ad-blockers is evolving.

\begin{figure}
\begin{lstlisting}
//step 1: set timeout
var myVar = setInterval(function() {
  myFunc()
}, 2000);

function myFunc() {

  // step 2: condition check
  if (window.iExist === undefined ||
    (!$("#XUinXYCfBvqpyDHOrOAVClxoWJemrlPpfYCdWfiyAzNY").is(
      ":visible") && (($(".vip_052x003").height() < 100 && !$(
      "#vipchat").length) && $(".vip_09x827").height() < 25))) {

      //step 3: response
      $("#XUinXYCfBvqpyDHOrOAVClxoWJemrlPpfYCdWfiyAzNY").css(
         "width:100%;height:100%;position:fixed;z-index:999999;top:0");
      $("#XUinXYCfBvqpyDHOrOAVClxoWJemrlPpfYCdWfiyAzNY").show();
  }
  else if ($("#XUinXYCfBvqpyDHOrOAVClxoWJemrlPpfYCdWfiyAzNY").is(
      ":visible") && $(".vip_052x003").height() > 249) {

      $("#XUinXYCfBvqpyDHOrOAVClxoWJemrlPpfYCdWfiyAzNY").hide()
  }
}
\end{lstlisting}
\vspace{-0.1in}
\caption{Ad-block detection JavaScript extracted from \protect \url{http://www.vipleague.tv}}
\label{fig:adbcode}
\vspace{-0.1in}
\end{figure}

\vspace{0.05in} \noindent \textbf{Illustration of ad-block detection.}
To understand how ad-block detection scripts operate, let's analyze the complete cycle of ad-block detection.
Figure \ref{fig:adbtonoadb} shows the web page loading process of \url{http://www.vipleague.tv}, which employs ad-block detection, on a web browser with Adblock Plus.
Figure \ref{fig:adbcode} shows the JavaScript that is used for ad-block detection by the website.\footnote{At the time of writing, this script can detect users with Adblock Plus.}
The functionality of the JavaScript can be divided into three parts: timeout, condition check, and response.
In Figure \ref{fig:adbtonoadb}, we note that the web browser starts loading the HTML and other page content included in the HTML code (\ding{202}).
While the content is loading, ad-block extension kicks in and starts evaluating the HTML code and page content to remove potential ads (\ding{203}).
Since the ad-blocking software starts working after a small delay, the ad-block detection script has to wait some time before monitoring the ads.
In Figure \ref{fig:adbcode}, the timeout is set at 2000 milliseconds.
Once the timeout expires, the condition check is executed to verify the presence/absence of ads.
This step is typically carried out by accessing various elements and their \textit{css} properties.
In Figure \ref{fig:adbcode}, the script first checks the \textit{visibility} property of the div with identifier \texttt{XUinXYCfBvqpyDHOrOAVClxoWJemrlPpfYCdWfiyAzNY}, as will be discuss later, this div is specifically designed for ad-block detection.
Ad-blockers sometimes also make the ad invisible by decreasing its dimensions; therefore, the script verifies the height and length properties of the ad related div elements for classes \textit{vip\_052x003} and \textit{vip\_09x827}.
If the script detects that ads were removed or hidden, then the response step is executed.
As discussed earlier, the implementation details of this step varies across publishers.
A few publishers gently request users to remove/disable their ad-blockers, while others aggressively show a page-wide notification and/or block content (\ding{204}).
For example, in Figure \ref{fig:adbcode}, the publisher responds by first changing css properties of the div, which it verifies in a conditional check.
More specifically, the publisher sets the z-index of the div to make it a pop-up message.

\presec \section{Measuring Ad-block Detection} \postsec
\label{sec:measurement}
In this section, we design and implement our approach for automatically identifying websites that employ ad-block detection.
The main premise of our approach is that websites conducting ad-block detection make distinct changes to their web page content for ad-block users as compared to users without ad-block.
Our goal is to identify, quantify, and extract such distinct features that can be leveraged for training machine learning models to automatically detect websites that employ ad-block detection.

\subsection{Overview}
We want to identify distinct features that capture the changes made by ad-block detectors to the HTML structure of web pages.
To this end, we first conducted some pilot studies to test the behavior of websites that employ ad-block detection.
Based on our pilot studies, we found that the changes made by ad-block detectors can be categorized into: (1) addition of extra DOM nodes, (2) change in the style of existing DOM nodes, and (3) changes in the textual content.
We also found a few cases when the websites completely changed the web page content.
In addition, a few websites with ad-block detectors reacted by redirecting users to warning pages.
Note that the Adblock Plus is installed with the default configuration which allows acceptable ads~\cite{acceptable-ads}.
This will likely suppress many ad-block detections and result in underestimating their prevalence.
However, since most regular users would choose the default configuration, we believe our study represents what most users would observe regarding to ad-block detection.
Below, we provide an overview of our proposed features and also discuss how they capture the changes by ad-block detectors.

\vspace{0.05in} \noindent \textbf{Node additions.}
We found that in order to show notification to users with ad-blockers, websites dynamically create and add new DOM nodes.
Thus, node additions in the DOM can potentially indicate ad-block detection.
We can log the total number of DOM elements inserted in a web page.

\vspace{0.05in} \noindent \textbf{Style changes.}
We found that a few websites include ad-block detection notifications which are in their page content but hidden.
If these websites detect the use of ad-blockers, they change the visibility of their notification.
To cover such cases, we can log attribute changes to DOM elements of a web page.

\vspace{0.05in} \noindent \textbf{Text changes.}
Other then structural changes, we found that some websites change the textual content (i.e., text-related nodes) in response to ad-blockers.
Therefore, we can log changes in the textual content of a web page and addition of text-related nodes in a web page.

\vspace{0.05in} \noindent \textbf{Miscellaneous features.}
In addition to the above-mentioned features, we also consider other features like innerHTML to detect whether the structure is completely changed and URL to detect redirection.

\begin{figure}[!t]
\includegraphics[width=1\columnwidth]{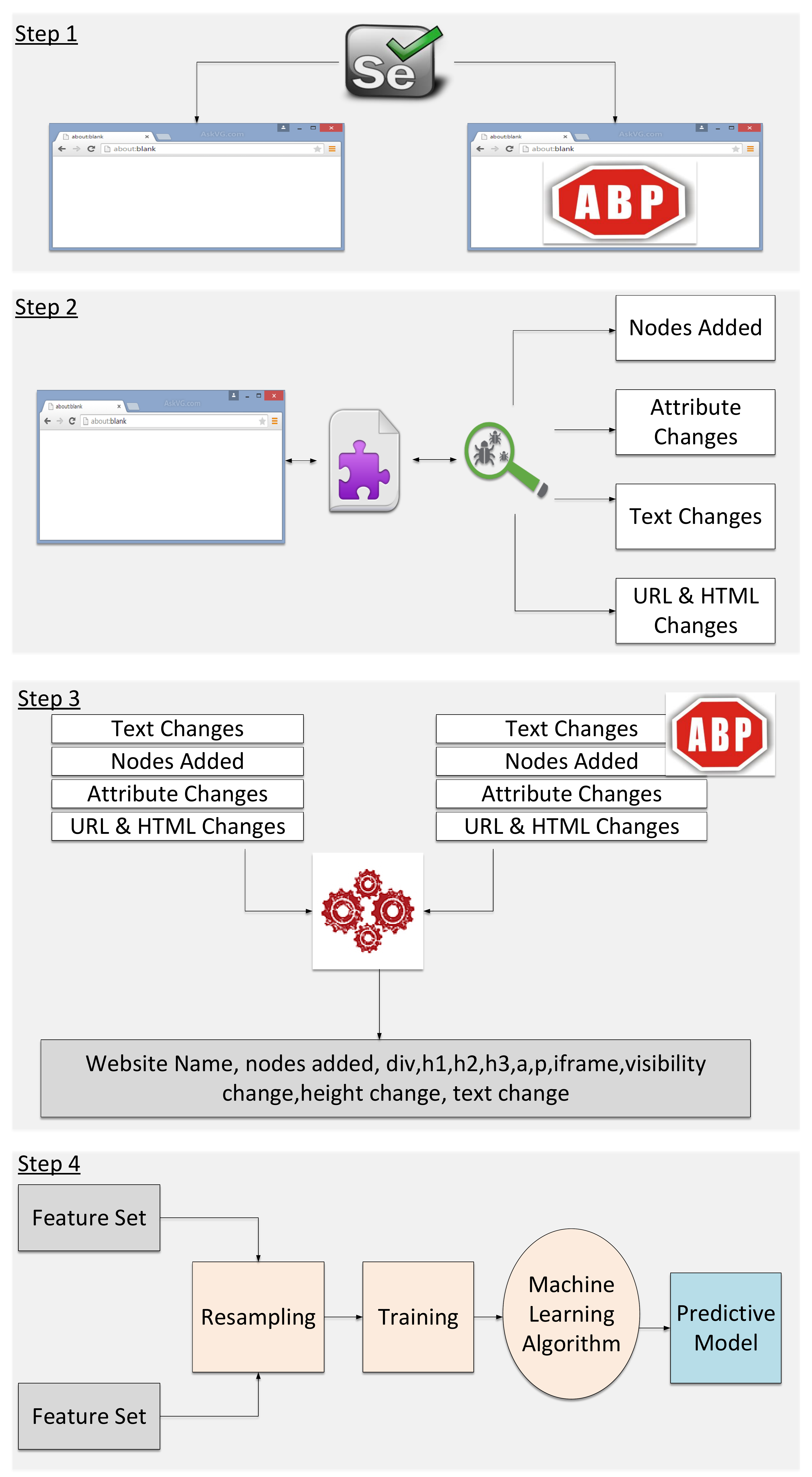}
\caption{Overview of our methodology for measuring ad-block detection.}
\label{fig:adb_implementation}
\end{figure}

\subsection{Methodology}
Figure \ref{fig:adb_implementation} provides an overview of our methodology to automatically measure ad-block detection on the web.
We conduct A/B testing to compare the contents of a web page with and without ad-blocking software.
To automate this process, we use the Selenium Web Driver \cite{selenium} to open two separate instances of the Chrome web browser, with and without Adblock Plus (\ding{202}).
We implemented a custom Chrome browser extension to record changes in the content of web pages during the page load process.
Our extension records the structure of the DOM tree, all textual content, and HTML code of the web page (\ding{203}).
We implemented a feature extraction script to process the collected data and generate a feature vector for each website (\ding{204}).
We feed the extracted features to a supervised classification algorithm for training and testing (\ding{205}).
We train the machine learning model using a labeled set of websites with and without ad-block detectors.
Below we describe these steps in detail.

\vspace{0.05in} \noindent \textbf{Web automation for A/B testing.}
Using the Selenium Web Driver \cite{selenium}, we implemented a web automation tool to conduct automated measurements.
For A/B testing, our tool first loads a website without Adblock Plus, and then opens it with Adblock Plus in a separate browser instance.
However, we found that many websites host dynamic content that changes at a very small timescales.
For example, some websites include dynamic images (e.g., logos), which can introduce noise in our A/B testing.
Similarly, most news websites update their content frequently which can also add noise.
Thus, we may incorrectly attribute these changes to the ad-blocker or ad-block detector used by the publisher.
To mitigate the impact of such noise, our tool opens multiple instances of each website in parallel and excludes content that changes across multiple instances.

\vspace{0.05in} \noindent \textbf{Data collection using a custom Chrome extension.}
To collect data while a web page is loading, we use DOM Mutation Observers \cite{mutation} to track changes in a DOM (e.g. \texttt{DOMNodeAdded}, \texttt{DOMAttrModified}, etc.).
The changes we track include addition of new DOM nodes or scripts, node attribute changes like class change or style change, removal of nodes, changes in text etc.
We implemented the data collection module as a Chrome extension.
The extension is preloaded in the browser instances that are launched by our web automation tool.
As soon as a web page starts loading, the extension attaches an observer listener with it.
Whenever an event occurs, the listener fires and we record the information.
For example, we record the identifier, type, value, name, parent nodes, and attributes of the corresponding node.
For each attribute change, in addition to above-mentioned information, we record the name of attribute which changes like style or class and its old and new value.
We also log page level data such as the complete DOM tree, innerText, and innerHTML as well.

\begin{table}[!t]
\centering
\caption{Features used to identify ad-block detectors}
\label{table:features}
\begin{tabular}{@{}ll@{}}
\rowcolor[HTML]{343434}
{\color[HTML]{FFFFFF} \textbf{Feature Set}}                  & {\color[HTML]{FFFFFF} \textbf{Description}} \\ 
\multicolumn{1}{l}{}                                           & {\# div}               \\ \cmidrule(l){2-2}
\multicolumn{1}{l}{}                                           & {\# h1}                \\ \cmidrule(l){2-2}
\multicolumn{1}{l}{}                                           & {\# h2}                \\ \cmidrule(l){2-2}
\multicolumn{1}{l}{}                                           & {\# h3}                \\ \cmidrule(l){2-2}
\multicolumn{1}{l}{}                                           & {\# img}               \\ \cmidrule(l){2-2}
\multicolumn{1}{l}{}                                           & {\# table}             \\ \cmidrule(l){2-2}
\multicolumn{1}{l}{}                                           & {\# p}                 \\ \cmidrule(l){2-2}
\multicolumn{1}{l}{}                                           & {\# br}                \\ \cmidrule(l){2-2}
\multicolumn{1}{l}{}                                           & {\# iframe}            \\ \cmidrule(l){2-2}
\multicolumn{-4}{l}{\multirow{-10}{*}{\textbf{Node features}}} & {\# nodes}       \\ \midrule
                                                               & {display change}              \\ \cmidrule(l){2-2}
                                                               & {visibility change}           \\ \cmidrule(l){2-2}
                                                               & {height change}               \\ \cmidrule(l){2-2}
                                                               & {width change}                \\ \cmidrule(l){2-2}
                                                               & {opacity change}              \\ \cmidrule(l){2-2}
                                                               & {maxheight change}            \\ \cmidrule(l){2-2}
                                                               & {background-size change}      \\ \cmidrule(l){2-2}
\multirow{-4}{*}{\textbf{Attribute features}}                  & {total changes in style}      \\ \midrule
                                                               & {lines differences}           \\ \cmidrule(l){2-2}
                                                               & {character differences}       \\ \cmidrule(l){2-2}
\multirow{-4}{*}{\textbf{Text features}}                       & {bag of words}                \\ \midrule
                                                               & {HTML changes}                \\ \cmidrule(l){2-2}
\multirow{-4}{*}{\textbf{Overall}}                            & {URL change}                  \\ \bottomrule
\end{tabular}
\vspace{-.3in}
\end{table}

\vspace{0.05in} \noindent \textbf{Feature extraction.}
We then process the output of data collector to extract a set of informative features which can distinguish between changes due to ad-block detection.
Recall that we load each page multiple times to mitigate noise.
Let \textbf{A} denote the data collected with ad-blocker, and let \textbf{B} \& \textbf{B'} denote the data collected by loading a web page twice without an ad-blocker.
We provide details of the feature extraction process below.
Table \ref{table:features} includes the list of all features used in our study.

\noindent $\bullet$ \textit{Node features.}
For each instance, we extract DOM related nodes because our pilot experiments revealed that websites using ad-block detection add only DOM related nodes.
More specifically, we extract the list of \texttt{anchor}, \texttt{div}, \texttt{h1}, \texttt{h2}, \texttt{h3}, \texttt{img}, \texttt{table}, \texttt{p}, and \texttt{iframe} nodes for each instance.
Once we have a list of DOM nodes for each instance, we compare \textbf{A} vs. \textbf{B'} and \textbf{B} vs. \textbf{B'} to obtain the list of differences between these nodes.
We denote these lists as \textbf{AB'} and \textbf{BB'} lists.
As explained earlier, to remove number of node differences due to dynamic content of websites, we cross-validate nodes in \textbf{AB'} with \textbf{BB'} using their properties.
Our key idea is that if a publisher ads random nodes to a web page, they may have different identifiers but most the other properties will be almost similar.
Thus, we remove the nodes from \textbf{AB'} that also appear in \textbf{BB'}.

\noindent $\bullet$ \textit{Attribute features.}
For each instance, we extract changes in the style of DOM related nodes.
More specifically, we focus on changes to the display-related property of nodes.
For instance, we log whether the visibility property of a node changes from hidden to non-hidden.
We also log changes to the display property of a node, e.g., the number of changes in height, width, and opacity of nodes.
Similar to node features, we compare \textbf{A}, \textbf{B}, and \textbf{B'} to eliminate attribute changes from \textbf{AB'} that also appear in \textbf{BB'}.

\noindent $\bullet$ \textit{Text features.}
We get the list of all text nodes in \textbf{A}, \textbf{B} and \textbf{B'}.
Using the lists, we identify pairs of nodes with differences texts.
We particularly focus on line differences rather than character-level differences to mitigate noise (e.g., difference in clock time).
We again compare \textbf{A}, \textbf{B}, and \textbf{B'} to eliminate changes in textual features from \textbf{AB'} that also appear in \textbf{BB'}.

\noindent $\bullet$ \textit{Structural features.}
We compare differences in the overall page HTML using the \emph{cosine similarity} metric.
If the cosine similarity between \textbf{A} and \textbf{B}/\textbf{B'} is very low, it indicates significant content change.
To check for potential URL redirections, we also track changes in URL.

\vspace{0.05in} \noindent \textbf{Classification model training and testing.}
We feed the extracted features to a machine learning classifier to automatically detect websites that employ ad-block detection.
However, in order to train the classification algorithm, we need a sufficient number of labeled examples of websites that detect ad-blockers (i.e., positive samples) and websites that do not detect ad-blockers (i.e., negative samples).
To get positive samples, we first use a crowd-sourced list of such websites \cite{antiadblockerkillerlist}.
We manually validated the websites in this list, and excluded websites that did not detect and respond to ad-blockers.
We also manually opened Alexa top 1000 websites and identified four websites that use ad-block detection.\footnote{
During the manual verification, we found that the response of websites after ad-block detection varies.
Most websites detect and respond to ad-blockers on the homepage without waiting for any input from users.
In contrast, some websites respond to ad-blockers only when a particular content type is requested (e.g., video is played) or when the user navigates to other pages.
Since it is not practical to automatically identify such requirements, we restrict ourselves to the former category of websites.
Also note that some websites include ad-block detection logic but they do not respond to ad-blockers.
We excluded these websites from the list as well.
}
Overall, we identified a total of 200 positive training samples.
Since a vast majority of Alexa top 1000 websites do not deploy ad-block detection, we use them as negative training samples.

\begin{table}[!t]
\caption{Feature ranking based on information gain}
\label{table:entropy}
\begin{tabular}{lr}
\hline
\rowcolor[HTML]{343434}

\multicolumn{1}{l}{\color[HTML]{FFFFFF} \textbf{Name}} & \multicolumn{1}{r}{\color[HTML]{FFFFFF} \textbf{Information Gain}} \\\hline

\textbf{\# words}   & 35.44\%     \\ \hline
\textbf{\# text nodes added}             & 27.89\%                                                                      \\ \hline
\textbf{\# lines added}                  & 18.13\%                                                                      \\ \hline
\textbf{\# nodes added}      & 17.37\%                                                                      \\ \hline
\textbf{\# characters added}             & 17.19\%                                                                      \\ \hline
\textbf{\# div nodes added}         & 13.01\%                                                                      \\ \hline
\textbf{\# height property changed}  & 10.67\%                                                                      \\ \hline
\textbf{\# display property changed}& 8.67\%                                                                      \\ \hline
\textbf{\# styles attribute changed}& 7.20\%                                                                      \\ \hline
\textbf{\# images added}           & 5.82\%                                                                      \\ \hline
\end{tabular}
\vspace{-.2in}
\end{table}

\begin{figure}[!t]
\centering
\subfigure{\includegraphics[width=0.75\columnwidth]{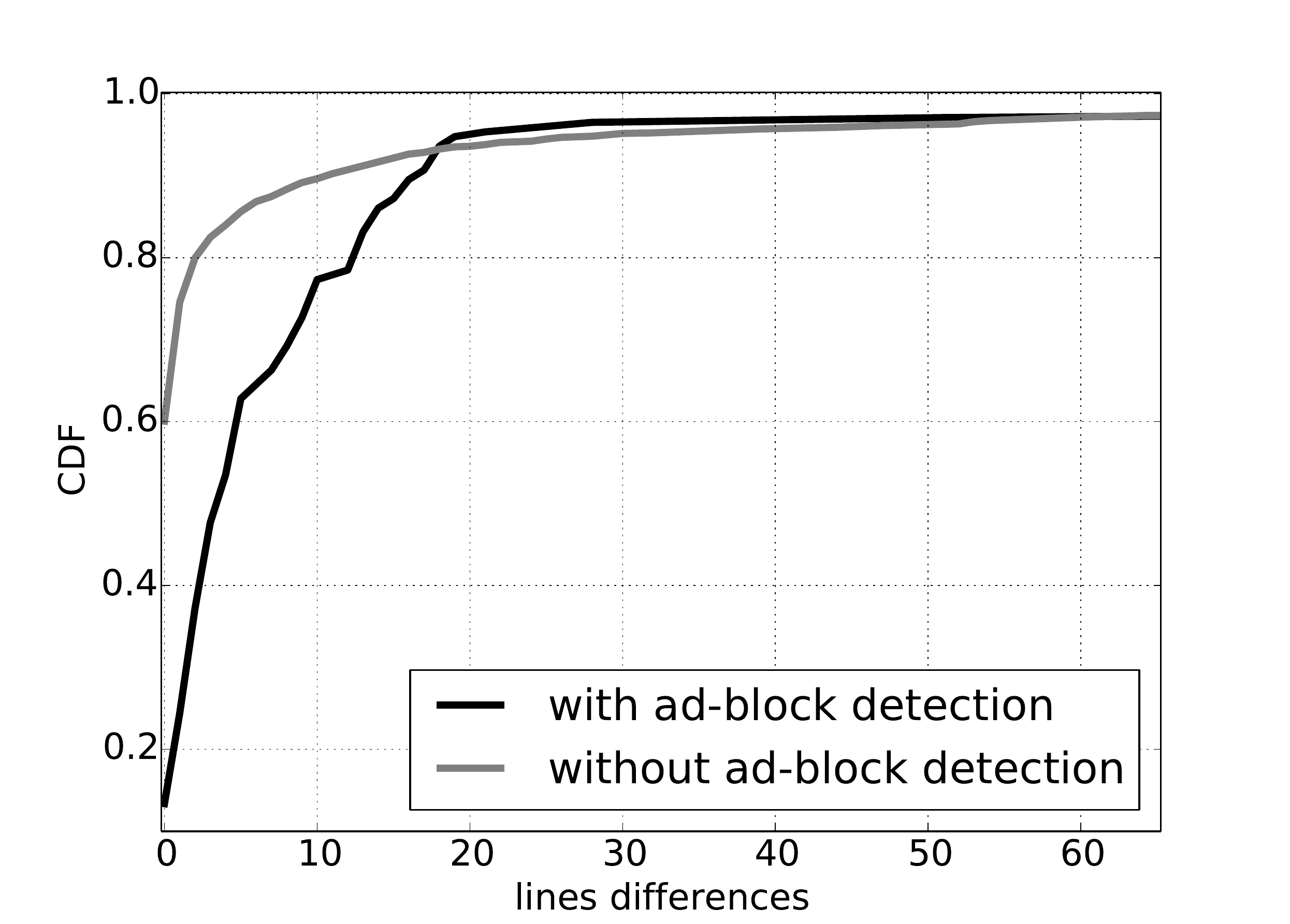}\label{fig:third_sub}}
\subfigure{\includegraphics[width=0.75\columnwidth]{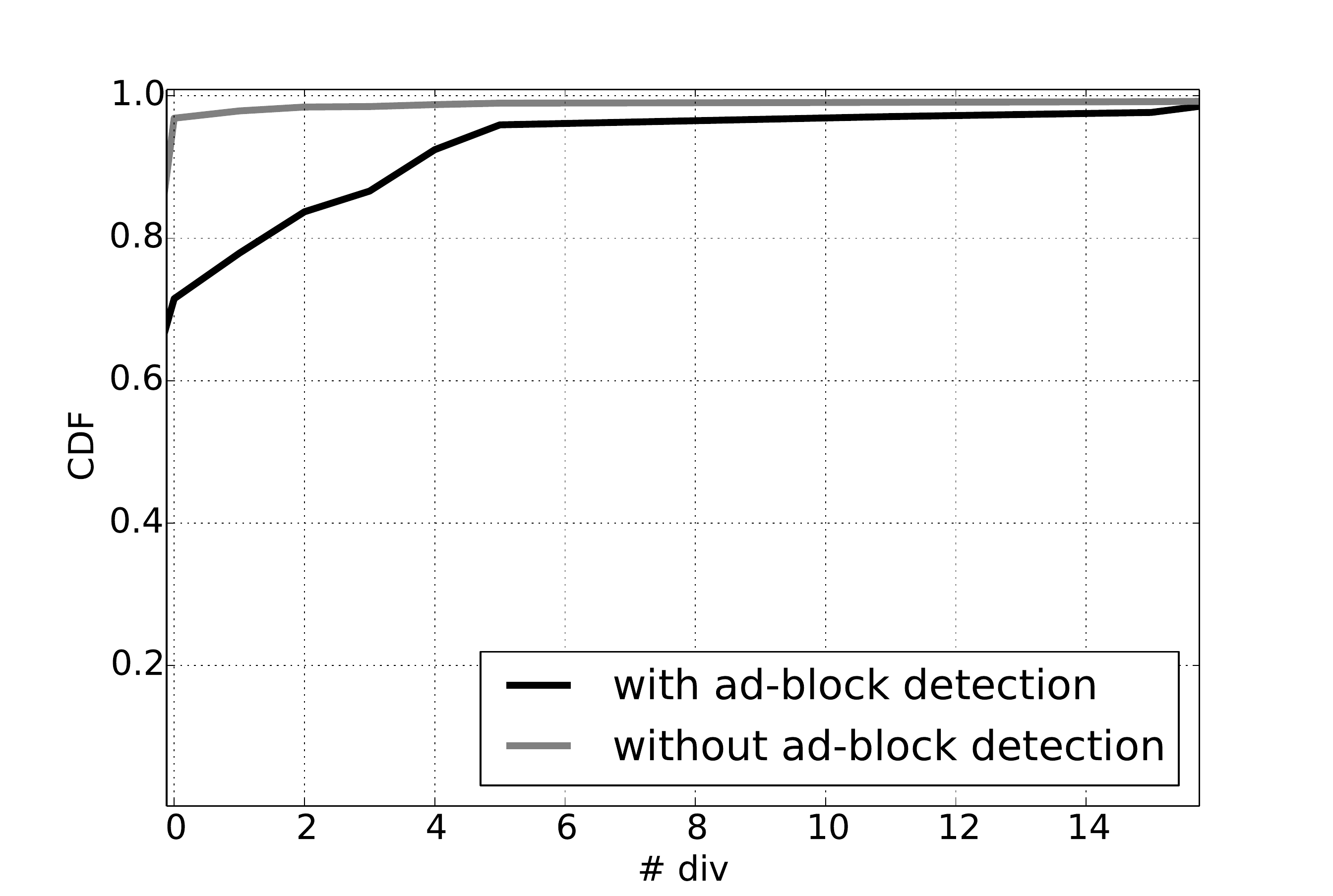}\label{fig:first_sub}}
\caption{Distribution of features used to identify ad-block detection}
\label{fig:graph_cdfs}
\vspace{-.2in}
\end{figure}

\subsection{Feature Analysis}
In this section, we analyze the extracted features to quantitatively understand their usefulness in identifying ad-block detection.
We first visualize the distributions of a few features.
Figure \ref{fig:graph_cdfs} plots the cumulative distribution functions (CDF) of two features.
We observe that websites which employ ad-block detection tend to changes more lines and add \texttt{div} elements than other websites.
%
%
These distributions confirm our intuition that ad-block detectors make changes in the web content that are distinguishable.

To systematically study the usefulness of different features, we employ the concept of \emph{information gain} \cite{mitchell97MLbook}, which uses entropy to quantify how our knowledge of a feature reduces the uncertainty in the class variable.
The key benefit of information gain over other correlation-based analysis methods is that it can capture non-monotone dependencies.
Let $H(X)$ denote the entropy (i.e., uncertainty) of feature $X$.
H is defined as:
\[
H=-\sum\limits_{i}{p_{i}{\log{p_{i}}}}
\]
Let $H(Y)$ denote the entropy (i.e., uncertainty) of the binary class variable $Y$.
Information gain is computed as:
\[
IG(Y|X) = H(Y) - H(Y|X).
\]
We can normalize information gain, also called \emph{relative information gain}, as:
\[
\frac{H(Y) - H(Y|X)}{H(Y)}.
\]
Using this, we can quantify what an input feature informs us about the use of ad-block detection.
Table \ref{table:entropy} ranks the top 10 features based on their information gain.
We note that text-based features (number of words changed and number of text nodes added) have the highest information gain, both exceeding 25\%.
They are followed by node and style based features (e.g., number of \texttt{div} elements added, number of nodes for which height property is changed, etc.).

\begin{table}[!t]
\centering
\caption{Effectiveness of different classifiers}
\label{table:classifiers}
\begin{tabular}{@{}lrrr@{}}
\rowcolor[HTML]{343434}
{\color[HTML]{FFFFFF} \textbf{Classifier}} & {\color[HTML]{FFFFFF} \textbf{Recall}} & {\color[HTML]{FFFFFF} \textbf{Precision}} & {\color[HTML]{FFFFFF} \textbf{AUC}} \\ 
\textbf{Random Forest}                     & \textbf{93.1\%}              & \textbf{94.8\%}     & \textbf{96.0\%}  \\ \midrule
\textbf{C4.5 Decision Tree}                & {87.0\%}     & {89.0\%}          & {91.3\%}                           \\ \midrule
\textbf{Naive Bayes}                       & {82.0\%}    & {82.4\%}                            & {89.0\%}          \\ \bottomrule
\end{tabular}
\end{table}


\subsection{Classifier Evaluation}
We train machine learning classification models using the labeled set of 1000 negative samples and 200 positives samples.
We use the standard $k$-fold cross validation methodology to verify the accuracy of the trained models.
For this purpose we select $k=5$, divide the data into 5 folds where one fold is used as training set while rest of folds are used for verification.
To quantify the classification accuracy of the trained models, we use the standard ROC metrics such as precision, recall, and area under ROC curve (AUC).

\[
\text{Precision} = \frac{\text{True Positives}}{\text{True Positives} + \text{False Positives}}
\]

\[
\text{Recall} = \frac{\text{True Positives}}{\text{True Positives} + \text{False Negatives}}
\]
We test multiple machine learning models on our data set.
We tuned various parameters of each of these models to optimize their classification performance.
Table \ref{table:classifiers} summarizes the classification accuracy of these classifiers.
We note that the random forest classifier, which is a combination of tree classifiers, clearly outperforms the C4.5 decision tree and the naive Bayes classifiers.
The random forest classifier achieves 93.1\% recall, 94.8\% precision, and 96.0\% AUC.

To further evaluate the effectiveness of different feature sets in identifying ad-block detection, we conduct experiments using stand alone feature sets and
then evaluate their all possible combinations.
We divide the features into node features, attribute features, and text features.
Among stand alone feature sets, text-based features provide the best classification accuracy.
We also observe that using combinations of feature sets does improve the classification accuracy.
The best classification performance is achieved when all feature sets are combined.

\begin{figure}[!t]
\includegraphics[width=1\columnwidth]{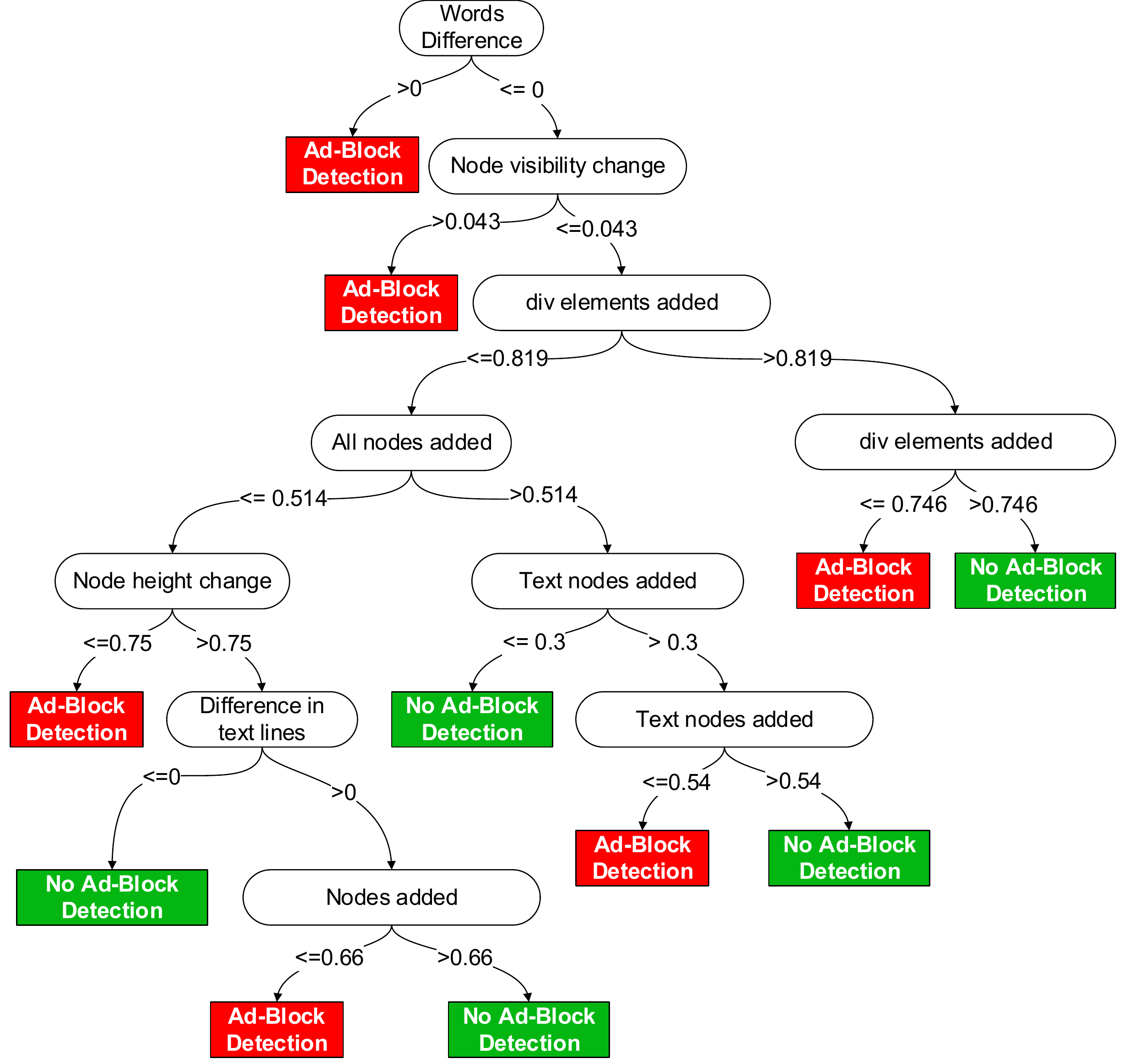}
\caption{Visualization of decision tree model for ad-block detection}
\label{fig:tree_decisiontree}
\end{figure}

To further gain some intuition from the trained machine learning models, we visualize a pruned version of the decision tree model trained on labeled data in Figure \ref{fig:tree_decisiontree}.
As expected from the information gain analysis, we note that a text feature (\emph{words difference}) is the root node of the decision tree.
If there is a positive word difference, the model detects ad-block detection.
Similarly, if node visibility is changed, the model detects ad-block detection.
It is interesting to note that the top three features in the decision tree belong to different feature categories.
This indicates that different feature sets complement each other, rather than capturing similar information, which we also observed earlier when evaluating different combinations of features.

\section{Ad-block Detection in the Wild}  \label{sec:adblock_measurement}
We want to analyze the strategies and methods used by publishers for ad-block detection.
To this end, we first use the random forest model on Alexa top 100K websites to identify ad-block detectors.
Our machine learning model found a total of 292 websites that detect and respond to ad-blockers.
Table \ref{my-label} (in Appendix) lists these 292 ad-block detecting websites along with their Alexa rank.
We note that a vast majority of the websites in Table \ref{my-label} have low Alexa ranks, likely due to
1) the top web websites have paid ad-blockers to be whitelisted or 2) the top websites are worried about
losing users if they take an aggressive stance against ad-blocker users.
Using additional string based features (e.g., ``Adblock'', ``Adblock Plus''), we also found a total of 797 websites that have ad-block detection scripts but do not exhibit visible behaviors, likely due to default-on acceptable ads in our Adblock Plus extension.
It is also possible that such websites are currently tracking the usage of ad-blockers
but not necessarily ready to go aggressively against users.
Overall, we found 1,089 ad-block detecting websites in the Alexa top-100K list.
In this section, we focus our attention on the ad-block detecting websites that not only detect ad-blockers but also respond to them.

Our goal here is to characterize \emph{how} different ad-block detection strategies operate under the hood.
We cluster ad-block detection strategies based on their JavaScript code similarity.
Our analysis allows us to measure the popularity of specific strategies and third-party ad-block detection services, \eg PageFair.
The result of the analysis will also help us design countermeasures against the state-of-the-art ad-block detectors.



\subsection{JavaScript Collection}
As a first step, we collect the JavaScript code of all websites that employ ad-block detection.
Analyzing the functionality of JavaScript code is non-trivial because the code can be packed inside functions such as \texttt{eval}.
To overcome these issues, we leverage the fact that the code needs to unpack itself before execution.
We attach a debugger between the Chrome V8 JavaScript engine \cite{v8engine} and the web pages.
Specifically, we observe \texttt{script.parsed} function, which is invoked when \texttt{eval} is called or new code is added with \texttt{<iframe>} or \texttt{<script>} tags.
We implement the debugger as a Chrome extension and collect all JavaScript snippets parsed on a web page and identify the snippet responsible for ad-block detection.

\begin{figure*}[ht!]
    \subfigure
    {
        \includegraphics[width=1\columnwidth]{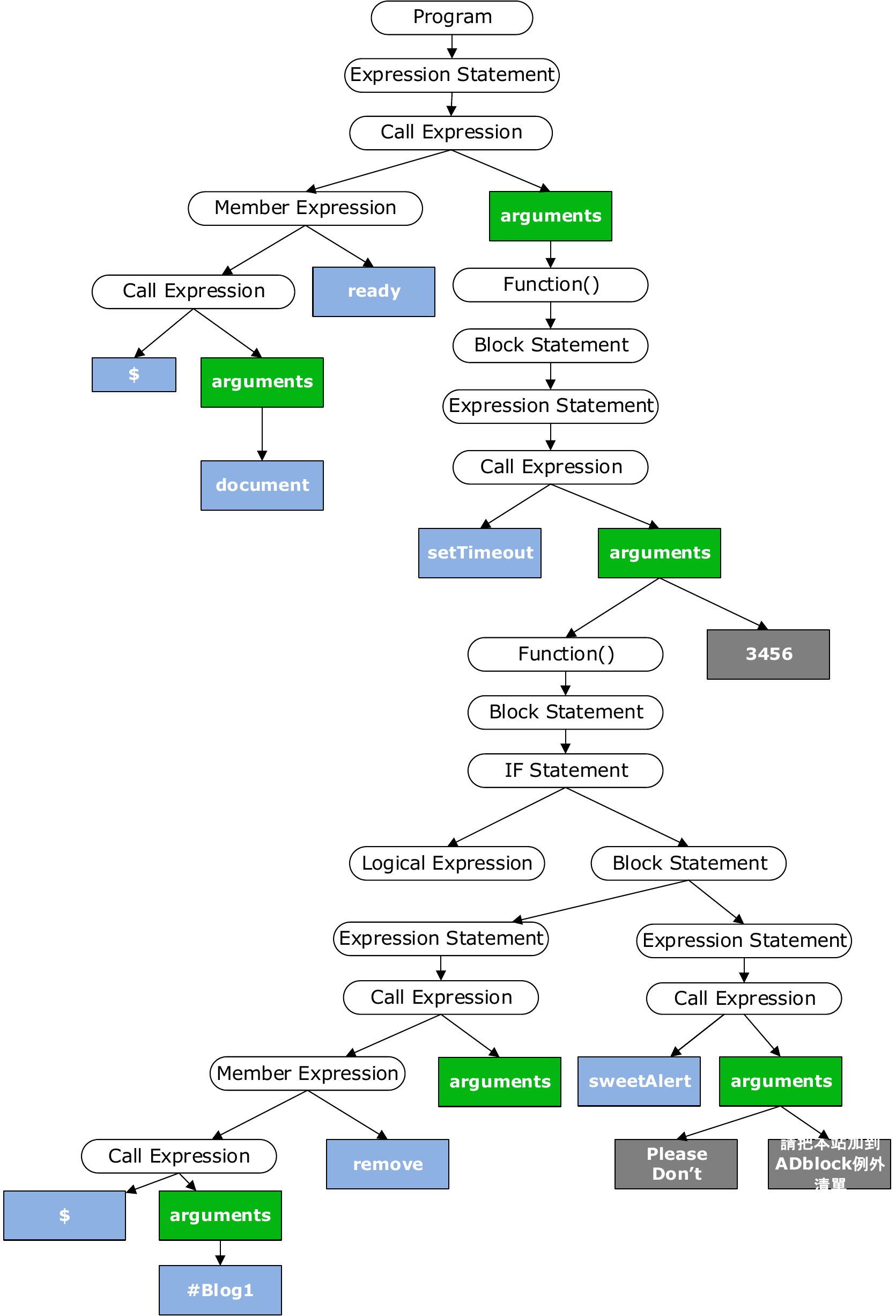}
        \label{fig:tree_ast1}
    }
    \subfigure
    {
        \includegraphics[width=1\columnwidth]{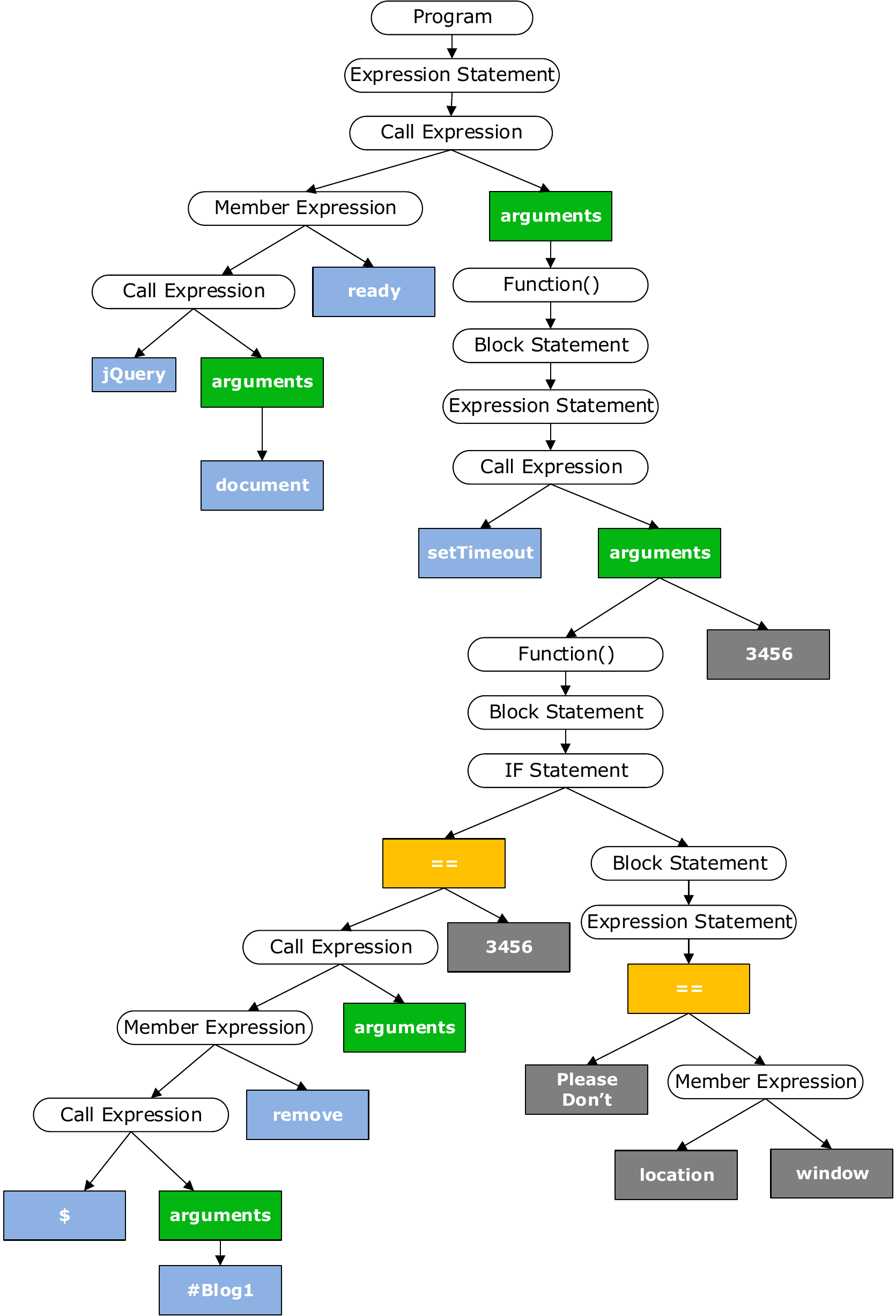}
        \label{fig:tree_ast2}
    }
   \begin{minipage}[\vtop]{1\columnwidth}
\begin{lstlisting}
$(document).ready(function() {
    setTimeout(function() {
        if (localStorage.noad === undefined && (16 >= $("#gAds").height() ||
          16 >= $("#gAd2".height())) {
            $("#Blog1").remove();
            sweetAlert("Oops.. please don't block my ADs",
                "warning");}
    }, 3456)
});
\end{lstlisting}
\end{minipage}
\begin{minipage}[\vtop]{1\columnwidth}
\vspace{-2.4cm}
\begin{lstlisting}
jQuery(document).ready(function() {
  setTimeout( if (jQuery("#adblock").height() == 0) {
                    window.location = "/adblock"
                }, 3456)
\end{lstlisting}
\end{minipage}
\vspace{-0.1in}
\caption{Visualization of ASTs of two ad-block detector JavaScript snippets. We note that although the code snippets appear to be different, the structure of their ASTs are similar.}
\label{fig:astCompare}
\vspace{-0.2in}
\end{figure*}

\subsection{Clustering}
Given these ad-block detector JavaScript snippets, we aim to cluster them into a few groups to identify their families.
To this end, we first compute ``similarity'' between JavaScript snippets and then use clustering.

\vspace{0.05in} \noindent \textbf{Methodology.}
To analyze and quantify the similarity between JavaScript snippets, we parse them to produce abstract syntax trees (ASTs).
ASTs have been used in prior literature for JavaScript malware detection \cite{curtsinger11zoozle,kapravelos2013revolver}.
ASTs allow us to retain the structural and logical properties of the code while ignoring fine details like variable names, which are not useful for our analysis.
Figure \ref{fig:astCompare} shows two ad-block detection JavaScript snippets and their corresponding AST visualizations.
We use the Esprima JavaScript parser \cite{esprima} to visualize ASTs for each JavaScript snippet.
We note that although the JavaScript snippets look fairly different but their ASTs have similar logical structure except minor differences near the leaf nodes.

We transform ASTs of all ad-block detection JavaScript snippets to normalized node sequences by performing the pre-order traversal on each tree.
Each variable length sequence is composed of node types that appear in the tree.
Note that there are 88 distinct node types in the JavaScript language.
To transform the variable length normalized node sequences to a fixed number of dimensions, we convert each sequence into a 88-dimensional summary vector.
Each JavaScript snippet is represented as an 88-dimensional point, where each dimension corresponds to a node type.
The value of each dimension is the node type frequency.

\begin{figure}[!t]
\centering
\includegraphics[width=0.7\columnwidth]{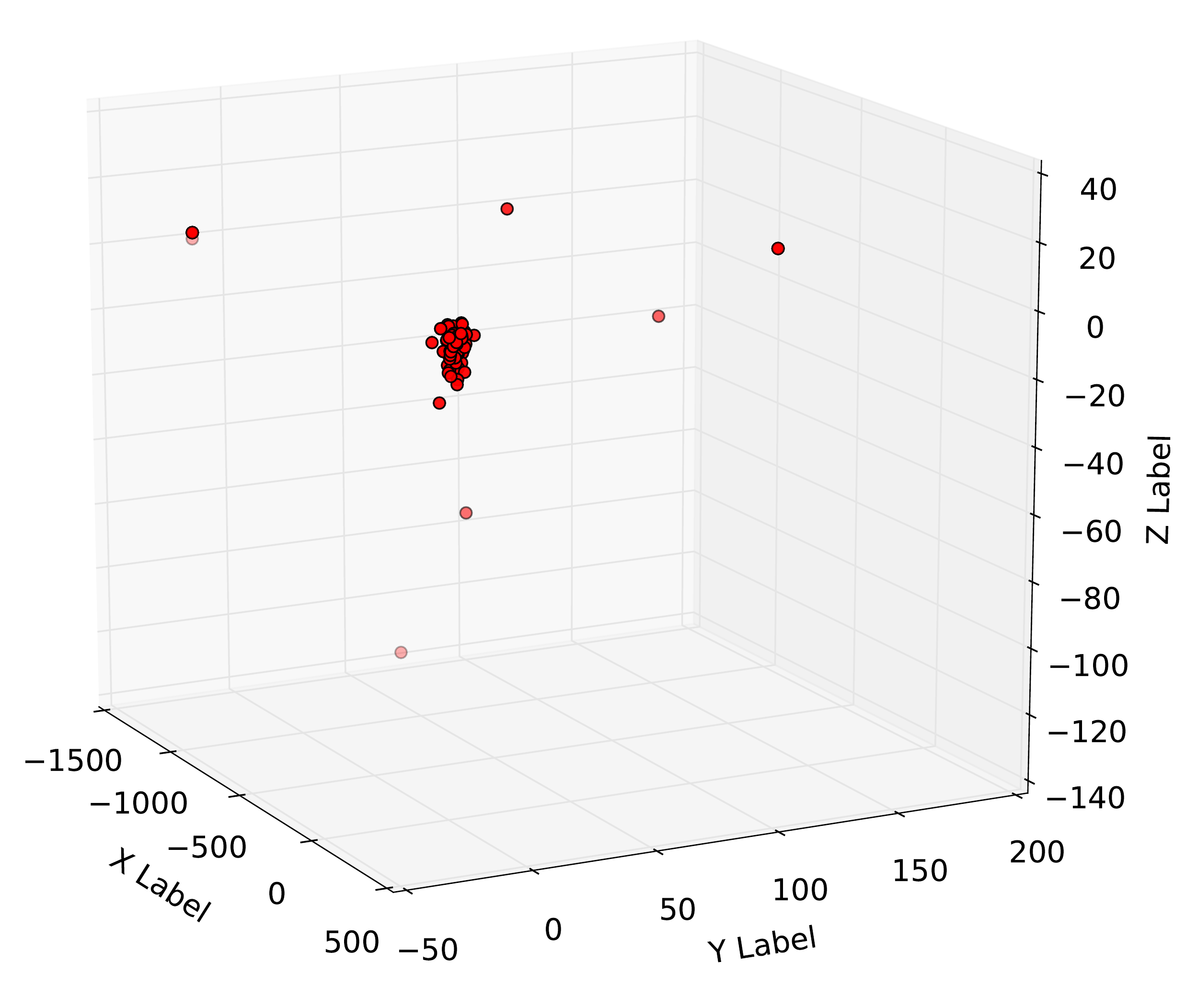}
\caption{Cluster visualization using PCA. The websites in the dense central cluster use simple scripts for ad-block detection. The outliers represent websites that use more sophisticated third-party scripts for ad-block detection. }
\label{fig:graph_pca}
\end{figure}

\vspace{0.05in} \noindent \textbf{Results.}
We use the Principal Component Analysis (PCA) to reduce the dimensionality of the summary vector for visualization.
Figure \ref{fig:graph_pca} plots the a 3-dimensional visualization of 292 ad-block detection JavaScript snippets.
We note that the center of the plot contains a dense cluster of instances.
Other outlier instances are spread out far from the central cluster.
We surmise that the central cluster represents websites that use a similar approach towards ad-block detection.
However, there are a number of outliers that represent customized and potentially more sophisticated approaches.
In the next section, we conduct an in-depth analysis of various ad-block detection approaches.

\subsection{Case Studies}
Next, we analyze the ad-block detection strategies used by different clusters.
We first study the ad-block detection strategies of websites in the dense central cluster.
We refer to these websites as the ``common family''.
We then study the ad-block detection strategies of outlier websites.
Our manual inspection of outliers revealed that these approaches use third-party ad-block detection scripts, including PageFair~\cite{pagefair}, FuckAdBlock~\cite{fuckadblocker}, and Sourcepoint~\cite{sourcepoint}.
Below, we provide an in-depth analysis of both types of websites.

\subsubsection{The Common Family}
The most distinct feature of ad-block detection JavaScript snippets in the common family is their simplicity.
Most of them are between 5-10 lines of code, yet they can successfully detect state-of-the-art ad-blockers.
Specifically, as discussed in \S\ref{sec:background}, ad-blockers tend to aggressively block ads by removing the ad frames entirely, without the intention of hiding their operation whatsoever.
The obvious nature of ad-blockers allows simple scripts, such as those in the common family, to easily identify ad-block users.

\vspace{0.05in} \noindent \textbf{Detection timing.}
We note that all websites in the common family launch their ad-block detection logic in the beginning of the page load process.
Since it may take a few seconds before an ad-blocker can remove the ads, some websites delay the execution of their logic by standard \texttt{setTimeout()} or \texttt{setTimeIntervel()}.
In Figure~\ref{fig:astCompare}, we show two example ad-block detection scripts, one inserting delay and another without delay.
Since the ad-block detection logic is a one-time check (i.e., it is not invoked periodically), ad-block detectors include the delay to ensure that ad-blockers have ample time conduct their operation.

\vspace{0.05in} \noindent \textbf{Detection logic.}
The ad-block detection JavaScript snippets in the common family typically check different ad elements to detect ad-blockers.
In Figure \ref{fig:validation}, we show the detection logic implemented by several websites in the common family.
We again note that the detection checks are fairly intuitive and simple.
For example, consider \texttt{urlchecker.org}, which checks whether the \textit{height} of \textit{adcheker} \texttt{div} is less then 10pxs or not.
Our further analysis revealed that the ad-blocker blocks the \textit{adsbygoogle.js} script due to which the \textit{adcheker} \texttt{div} is empty and its height is equal to 1px.
Other websites in Figure \ref{fig:validation} also check the properties of different \texttt{div} elements.
Since the filter lists used by ad-blockers, e.g., EasyList \cite{easylist} and Fanboy \cite{fanboy}, are publicly available, ad-block detectors can successfully setup these detection rules.
For instance, EasyList \cite{easylist} used by ad-blockers has \emph{adsbyggole.js} in its block list.

\vspace{0.05in} \noindent \textbf{Response.}
Although the detection logic used by websites in the common family is similar, their response to ad-blockers vary widely.
Figure \ref{fig:reaction} lists a few of the responses.
For \textit{hentai.to}, a \textit{<p>} element requests users to disable the ad-blocker.
Since the original content is preserved, this approach is not aggressive.
However, for \textit{knowlet3389.blogspot.hk}, the \textit{\#Blog1} \texttt{div} is removed upon ad-block detection, which indicates that the website hides its content from ad-block users.
\textit{elahmad.com} also aggressive responds by redirecting ad-block users to a warning page.
Overall, we find a wide spectrum of responses to ad-block detection, ranging from gentle request messages to more aggressive redirection.

\begin{figure}
\begin{lstlisting}
//http://knowlet3389.blogspot.hk/
if (localStorage.noad === undefined && (16 >= $("#gAds").height()
|| 16 >= $("#gAd2").height()))

//http://www.elahmad.com/
if(jQuery("#adblock").height()==0)

//http://urlchecker.org/
if ($("#adchecker").height() < 10)

//http://www.hentai.to/
if(document.getElementById("tester")!=undefined)

//http://forum.pac-rom.com/
if(!ad || ad.innerHTML.length == 0 || ad.clientHeight === 0)


\end{lstlisting}
\vspace{-0.15in}
\caption{Examples of detection logic used by different ad-block detectors}
\label{fig:validation}
\vspace{-0.15in}
\end{figure}


\subsubsection{PageFair}
PageFair is a service that allows publishers to detect ad-block usage and take mitigation actions such as display tailored non-intrusive ads to ad-block users. %
In the list of 292 ad-block detection websites, we found 12 websites that use PageFair.
We note that PageFair uses dynamic JavaScript and code obfuscation.
For example, it involves assembling URLs on the fly to retrieve additional JavaScript codes.
%
%
Below, we shed light on PageFair's strategy for ad-block detection and its response.

\begin{figure}
\begin{lstlisting}
//http://knowlet3389.blogspot.hk/
$("#Blog1").remove();
sweetAlert("Oops.. please don't block my ADs",
  "warning");

//http://www.elahmad.com/
window.location="/adblock"

//http://urlchecker.org/
$("#ads_notify").fadeIn();
$("#getlinks").hide();
$("#adchecker_btn").fadeIn();

//http://www.hentai.to/
document.write(
  '<p class="no">Please <u>disable</u> your adblocking
  software on hentai. TO to keep our community <u>FREE</u>! ^.^</p>'
);

//http://forum.pac-rom.com/
 alert(
   "We've detected an ad blocker running on your browser..."
 );


\end{lstlisting}
\vspace{-0.15in}
\caption{Examples of responses by different ad-block detectors.}
\label{fig:reaction}
\vspace{-0.2in}
\end{figure}

\vspace{0.05in} \noindent \textbf{Detection timing.}
PageFair performs multiple periodic checks at various stages of the web page load process to detect ad-blockers.
This approach is much more sophisticated than the common family and makes it harder for ad-blockers to evade detection by simply delaying their activity.

\vspace{0.05in} \noindent \textbf{Detection logic.}
PageFair's detection logic attempts to actively trap ad-blockers by injecting different ``baits'' on web pages.
This is in stark contrast to simple passive detectors used by most websites in the common family.
In addition, PageFair attempts to check whether any ad-block plug-in is installed by looking for various browser resources exposed by ad-blocking extensions.
The use of these methods makes PageFair's detection logic difficult to evade.
We separately discuss both of these methods below.

$\bullet$ \textit{Baiting:}
Figure \ref{fig:fakeads} shows different types of baits used by PageFair.
The first example shows an injected \texttt{div} element that is not visible on the page, \ie 1x1 in size and negative values of the \textit{top} and \textit{left} properties.
The other two examples show \texttt{img} and \texttt{script} baits.
PageFair's detection logic injects these baits with keywords such as ``ad'' in the element name or URL.
For example, the identifier of the \texttt{div} element is set to \textit{influads\_block}.
Similarly, the source of \texttt{script} tag is set to \emph{adsense.js}, which is a common script used by Google Ads.

$\bullet$ \textit{Extension resources:}
For Chrome browser, we note that PageFair attempts to detect the presence of ad-blockers by accessing extension resources exposed by various ad-blockers at \texttt{chrome-extension://}.
Figure \ref{fig:hacking} shows how PageFair accesses extension resources to identify 8 popular ad-blockers including AdBlock, Adblock Plus, AdBlock Pro, AdBlock Premium, Adblock Super, Adguard, Ad Remover, and uBlock.
For each type of ad-blocker, it includes a unique extension identifier, \eg \texttt{gighmmpiobklfepjocnamgkkbiglidom} for AdBlock, and the resource file path.
Note that Chrome generally does not allow web pages to directly access extension resources unless an extension specifies resources as \texttt{web\_accessible\_} \texttt{resources} in the manifest file and makes them publicly accessible.
We find that these resources of various ad-blockers requested by PageFair are indeed publicly accessible.
Thus, ad-blockers are leaking the proof of their presence to ad-block detectors.
In addition, Figure \ref{fig:hacking} shows that critical resources such as whitelisted pages are also accessible.
For example, Adblock Plus exposes \emph{block.html}, which allows websites to get a list of blocked URLs.

\cut{
\begin{figure}
\begin{lstlisting}
//self calling initialization code
(function() {

  function async_load(script_url){
    var protocol = ('https:' == document.location.protocol ? 'https://' : 'http://');
    var s = document.createElement('script'); s.src = protocol + script_url;
    var x = document.getElementsByTagName('script')[0]; x.parentNode.insertBefore(s, x);
  }

  //reference code for each website
  bm_website_code = 'Unique Code';

  //scripts dynamically loaded  at document ready
  jQuery(document).ready(function(){async_load('asset.pagefair.com/measure.min.js')});
  jQuery(document).ready(function(){async_load('asset.pagefair.net/ads.min.js')});
})();

\end{lstlisting}
\vspace{-0.1in}
\caption{PageFair uses different baits to detect ad-blockers.}
\label{fig:initialization}
\vspace{-0.2in}
\end{figure}
}

\begin{figure}
\begin{lstlisting}
//DIV bait
 var b = document.createElement("DIV");
 // d = "influads_block"
 b.id = d;
 //c=''
 b.className = c;
 //1x1 div
 b.style.width = "1px";
 b.style.height = "1px";
 //div not located in visible frame
 b.style.top ="-1000px";
 b.style.left ="-1000px";
 document.body.appendChild(b);
 //jquery selector of created div
 c = jQuery("#" + d);
 // check if this div is hidden by ad-blocker
 d = c.is(":hidden") ? 1 : 0;
 //removing the created div
 c.remove();

//IMG bait
 f = document.createElement("IMG");
 //d="06db9294"
 f.id = d;
 f.style.width = "1px";
 f.style.height = "1px";
 f.style.top = "-1000px";
 f.style.left = "-1000px";
 document.body.appendChild(f);
 //c="http://asset.pagefair.com/adimages/textlink-ads.jpg"
 f.src = c

//SCRIPT bait
 b = document.createElement("SCRIPT");
 //d="295f89b1"
 b.id = d;
 b.type = "text/javascript";
 document.getElementsByTagName("head")[0].appendChild(b);
 //a= "http://asset.pagefair.com/adimages/adsense.js"
 b.src = a

\end{lstlisting}
\vspace{-0.15in}
\caption{PageFair uses different baits to detect ad-blockers.}
\label{fig:fakeads}
\vspace{-0.2in}
\end{figure}

\begin{figure}
\begin{lstlisting}
//accessible resources of ad-blocking extensions
var c = {
    //chrome internal ext domain:// folder name of each extension/ subfolder and name of resource
    adblock: "chrome-extension://gighmmpiobklfepjocnamgkkbiglidom/img/icon24.png",
    adblock_plus: "chrome-extension://cfhdojbkjhnklbpkdaibdccddilifddb/block.html",
    adblock_pro: "chrome-extension://ocifcklkibdehekfnmflempfgjhbedch/components/block/block.html",
    adblock_premium: "chrome-extension://fndlhnanhedoklpdaacidomdnplcjcpj/img/icon24.png",
    adblock_super: "chrome-extension://knebimhcckndhiglamoabbnifdkijidd/widgets/block/block.html",
    adguard: "chrome-extension://bgnkhhnnamicmpeenaelnjfhikgbkllg/elemhidehit.png",
    adremover: "chrome-extension://mcefmojpghnaceadnghednjhbmphipkb/img/icon24.png",
    ublock: "chrome-extension://epcnnfbjfcgphgdmggkamkmgojdagdnn/document-blocked.html"
  }

\end{lstlisting}
\vspace{-0.1in}
\caption{PageFair accesses extension resources to detect ad-blockers.}
\label{fig:hacking}
\vspace{-0.2in}
\end{figure}


\vspace{0.05in} \noindent \textbf{Response.}
PageFair provides a whitelist ad service under the acceptable ads manifesto \cite{acceptable-ads}.
To understand PageFair's service, we installed PageFair on a test website that uses Google Ads.
With ad-blocker, as expected, we find that the original ad is not available.
PageFair includes a replacement ad, which is not hosted on Google's domain.
Instead, the replacement ad is hosted on PageFair's domain --- \textit{adsfeed.pagefair.com}.
%
We found that PageFair's domain is whitelisted by EasyList \cite{easylist}, which is a default list on popular ad-blockers.
The following snippet is from EasyList \cite{easylist}:

\texttt{"\#\#.pagefair-acceptable"}

\texttt{"\textbar\textbar pagefair.net\textasciicircum\$third-party"}


The rule clearly indicates that PageFair is allowed to show acceptable ads on partner websites.
PageFair's response is reflective of the growing adoption of acceptable ads by many publishers and ad-blockers \cite{walls2015measuring}.

It is worth mentioning that when we rechecked some websites several weeks after we finished the initial experiments,
the behaviors of websites changed. For instance, \url{http://www.vipleague.tv} used to detect ad-blockers and simply
refuse to serve users via a large pop-up window. It is correctly classified as the common family.
However, in its recent version, it has evolved to behave similarly to PageFair. Instead of refusing the service,
the website switches the origin of the ads from \textit{adsrvmedia}, which is on the filter list of ad-blockers,
to the same local domain. Since the new ad URL is not on the filter list, the ad-blocker simply fails to
block it. This shows yet another new response that is similar in nature to PageFair.
%


\section{Related Work}
To the best of our knowledge, this paper presents the first large-scale measurement study of ad-block detection.
Since the arms race between publishers and ad-blockers is a recent phenomenon, prior work has mainly focused on understanding the dynamics of ads and ad-blockers.
Below, we discuss prior literature on online advertising and ad-blockers.

\vspace{0.05in} \noindent \textbf{Online Advertising.}
A large number of web publishers rely on online advertising.
However, online advertising has recently become more intrusive and annoying to end-uses.
Researchers have recently focused on various security and privacy aspects of online advertising.
Li et al. conducted the first large-scale study of malicious advertising (called \emph{malvertising}) on the web \cite{li2012knowing}.
Their analysis of 90,000 websites showed that not only malicious ads effect top websites but they also evade detection by various cloaking techniques.
Zarras et. al. also conducted a large-scale study to determine the extent at which user are exposed to malicious advertisements \cite{zarras14malads}.
Their measurement study of more then 60,000 ads showed that around 1\% ads exhibit malicious behavior.
They also showed that a few ad networks are more prone to supply malicious advertisements than others.

Other than malvertising, researchers have also analyzed privacy implications of online advertising.
Online advertising relies on sophisticated tracking of users across the web to target personalized ads.
Roesner et al. conducted a comprehensive active measurement study of third-party tracking on the web \cite{roesner12thirdpartytracking}.
They found more than 500 unique trackers on 1000 websites, with a few web trackers covering a large fraction of users' browsing activity.
Metwalley et al. conducted a passive measurement study to determine the extent of tracking on the web.
They found that more than 400 tracking services are contacted by the users and also 80\% of the users are tracked by at least one tracking service within a second after starting their surfing sessions \cite{metwalley2015online}.
Nath performed a measurement of 500K ad requests from 150K Android apps \cite{nath2015madscope}.
The analysis showed that although most ad networks collect targeting data on mobile apps, it does not significantly alter the way ads are chosen for many users.

\vspace{0.05in} \noindent \textbf{Ad-blockers.}
Due to the rapid increase in the use of ad-blockers, researchers are interested in measuring the use of ad-blockers.
Pujol et al. conducted a measurement study using passive network traces of thousands of users from a European ISP to quantify ad-block usage \cite{pujol15adblockinwild}.
Their results show that 22\% of users use AdBlock Plus.
They also found that ad-blocker users still generate significant ad traffic due their enrollment in the \emph{acceptable ads program}.
Walls et al. conducted a study of the whitelists used by ad-blockers for allowing acceptable ads \cite{walls2015measuring}.
They analyzed the evolution of ad-blocker whitelists and performed measurements on 8K websites.
Their analysis showed that whitelists contains around 5,936 filters and 3,545 unique publisher domains.
Their findings highlight that whitelists are inclined towards top ranked Alexa websites (59\% filters are for top 5000 websites).
Gugelmann et al. proposed a methodology to compliment manual filter lists of ad-blockers by automatically blacklisting intrusive ads \cite{gugelmann2015automated}.
They train a classifier on HTTP traffic statistics and identify around 200 new advertising and tracking services.

Since ad-block detection is a recent phenomenon, to the best of our knowledge, no prior work has studied the mechanisms used by ad-block detectors.
Our work aims to fill this gap by conducting a large-scale measurement of ad-block detection on the web.



\section{Conclusion}
We present the first large-scale study of ad-block detection on the web.
Our main observation is that about 1100 websites in the Alexa top-100K are currently performing ad-block detection.
Out of these, about 300 websites respond to ad-block detection by requesting users to disable ad-blockers through notifications to more aggressive blocking of website content.
We also find the increasing use of third party ad-block detectors such as PageFair.
We envision the arms race to continue in the coming years as we expect both ad-blockers and ad-block detectors to adapt in the future.


\begin{thebibliography}{10}

\bibitem{adp}
Adblock plus.
\newblock \url{https://adblockplus.org}.

\bibitem{antiadblockerkillerlist}
{Anti-Adblock Killer List}.
\newblock
  \url{https://github.com/reek/anti-adblock-killer/blob/master/anti-adblock-killer-filters.txt}.

\bibitem{diconnect}
Diconnect.me.
\newblock \url{https://disconnect.me}.

\bibitem{easylist}
Easy list.
\newblock \url{https://easylist-downloads.adblockplus.org/easylist.txt}.

\bibitem{easyprivacy}
Easy privacy.
\newblock \url{https://easylist-downloads.adblockplus.org/easyprivacy.txt}.

\bibitem{fanboy}
{Fanboy Annoyances List}.
\newblock \url{https://www.fanboy.co.nz}.

\bibitem{ghostery}
Ghostery.
\newblock \url{https://www.ghostery.com}.

\bibitem{selenium}
{Selenium WebDriver}.
\newblock \url{http://www.seleniumhq.org/projects/webdriver}.

\bibitem{mutation}
{W3C DOM4. W3C Recommendation 19 November 2015}.
\newblock http://www.w3.org/TR/dom/\#mutationobserver.

\bibitem{warning_removal_list}
Adblock warning removal list.
\newblock
  \url{https://easylist-downloads.adblockplus.org/antiadblockfilters.txt}, Nov
  2015.

\bibitem{acceptable-ads}
Allowing acceptable ads in adblock plus.
\newblock \url{https://adblockplus.org/acceptable-ads}, Nov 2015.

\bibitem{fuckadblocker}
Fuckadblock.
\newblock \url{https://github.com/sitexw/FuckAdBlock}, Nov 2015.

\bibitem{iabadreport}
{IAB internet advertising revenue report. 2014 full year results. An industry
  survey conducted by PwC and sponsored by the Interactive Advertising Bureau
  (IAB).}
\newblock
  \url{http://www.iab.com/wp-content/uploads/2015/05/IAB_Internet_Advertising_Revenue_FY_2014.pdf},
  April 2015.

\bibitem{esprima}
Javascript parser.
\newblock \url{http://esprima.org}, Nov 2015.

\bibitem{v8engine}
Open source v8 javascript engine.
\newblock \url{https://code.google.com/p/v8/}, Nov 2015.

\bibitem{pagefair}
Pagefair.
\newblock \url{https://pagefair.com/}, Nov 2015.

\bibitem{sourcepoint}
Sourcepoint.
\newblock \url{http://sourcepoint.com/}, Nov 2015.

\bibitem{yahoo_mail_incident}
{Yahoo Mail obstructing AdBlock users from signing in}.
\newblock \url{http://bit.ly/1TcgoaX}, Nov 2015.

\bibitem{curtsinger11zoozle}
C.~Curtsinger, B.~Livshits, B.~Zorn, and C.~Seifert.
\newblock {ZOZZLE: Fast and Precise In-Browser JavaScript Malware Detection}.
\newblock In {\em USENIX Security Symposium}, 2011.

\bibitem{bestadblockers}
C.~Fitchett.
\newblock {Best Free Ad Blocker Software}.
\newblock \url{http://bit.ly/1lr6sik}, April 2015.

\bibitem{goldstein2013cost}
D.~G. Goldstein, R.~P. McAfee, and S.~Suri.
\newblock {The Cost of Annoying Ads}.
\newblock In {\em WWW}, 2013.

\bibitem{gugelmann2015automated}
D.~Gugelmann, M.~Happe, B.~Ager, and V.~Lenders.
\newblock {An Automated Approach for Complementing Ad Blockers' Blacklists}.
\newblock {\em Privacy Enhancing Technologies (PETS)}, 2015(2):282--298, 2015.

\bibitem{guha2011privad}
S.~Guha, B.~Cheng, and P.~Francis.
\newblock Privad: Practical privacy in online advertising.
\newblock In {\em NSDI}, 2011.

\bibitem{kapravelos2013revolver}
A.~Kapravelos, Y.~Shoshitaishvili, M.~Cova, C.~Kruegel, and G.~Vigna.
\newblock {Revolver: An Automated Approach to the Detection of Evasive
  Web-based Malware}.
\newblock In {\em USENIX Security Symposium}, 2013.

\bibitem{appleadblock}
L.~Kelion.
\newblock {Apple brings ad-blocker extensions to Safari on iPhones}.
\newblock \url{http://www.bbc.com/news/technology-34173732}, September 2015.

\bibitem{li2012knowing}
Z.~Li, K.~Zhang, Y.~Xie, F.~Yu, and X.~Wang.
\newblock {Knowing Your Enemy: Understanding and Detecting Malicious Web
  Advertising}.
\newblock In {\em ACM CCS}, 2012.

\bibitem{metwalley2015online}
H.~Metwalley, S.~Traverso, M.~Mellia, S.~Miskovic, and M.~Baldi.
\newblock {The Online Tracking Horde: A View from Passive Measurements}.
\newblock In {\em Traffic Monitoring and Analysis}. 2015.

\bibitem{mitchell97MLbook}
T.~Mitchell.
\newblock {\em {Machine Learning}}.
\newblock Mc-Graw-Hill, 1997.

\bibitem{nath2015madscope}
S.~Nath.
\newblock {MAdScope: Characterizing Mobile In-App Targeted Ads}.
\newblock In {\em MobiSys}, pages 59--73. ACM, 2015.

\bibitem{pujol15adblockinwild}
E.~Pujol, O.~Hohlfeld, and A.~Feldmann.
\newblock {Annoyed Users: Ads and Ad-Block Usage in the Wild}.
\newblock In {\em ACM Internet Measurement Conference (IMC)}, 2015.

\bibitem{roesner12thirdpartytracking}
F.~Roesner, T.~Kohno, and D.~Wetherall.
\newblock {Detecting and Defending Against Third-Party Tracking on the Web}.
\newblock In {\em USENIX Symposium on Networked Systems Design and
  Implementation (NSDI)}, 2012.

\bibitem{pagefairreport}
T.~P. Team.
\newblock 2015 adblocking report.
\newblock \url{https://blog.pagefair.com/2015/ad-blocking-report}, 2015.

\bibitem{walls2015measuring}
R.~J. Walls, E.~D. Kilmer, N.~Lageman, and P.~D. McDaniel.
\newblock Measuring the impact and perception of acceptable advertisements.
\newblock In {\em ACM IMC}, 2015.

\bibitem{zarras14malads}
A.~Zarras, A.~Kapravelos, G.~Stringhini, T.~Holz, C.~Kruegel, and G.~Vigna.
\newblock {The Dark Alleys of Madison Avenue: Understanding Malicious
  Advertisements}.
\newblock In {\em ACM IMC}, 2014.

\end{thebibliography}
\end{document}